\documentclass[aps,prl,showpacs,superscriptaddress,twocolumn]{revtex4-1}

\usepackage{graphicx}
\usepackage{color}
\usepackage{amsmath}
\usepackage{float}
\usepackage{verbatim}
\usepackage{longtable}
\makeatletter
\newcommand{\rmnum}[1]{\it \romannumeral #1}
\newcommand{\Rmnum}[1]{\rm \expandafter\@slowromancap\romannumeral #1@}
\makeatletter 

\begin{document}

\title{SCAN+rVV10: A promising van der Waals density functional}

\author{Haowei Peng}
  	\email{Haowei.Peng@gmail.com. Please contact H.P.\,for the implementation of rVV10 within VASP.}
	\affiliation{Department of Physics, Temple University, Philadelphia, PA 19122, USA}
\author{Zeng-Hui Yang}
	\affiliation{Department of Physics, Temple University, Philadelphia, PA 19122, USA}
\author{Jianwei Sun}
	\affiliation{Department of Physics, Temple University, Philadelphia, PA 19122, USA}
\author{John P. Perdew}
	\affiliation{Department of Physics, Temple University, Philadelphia, PA 19122, USA}
	\affiliation{Department of Chemistry, Temple University, Philadelphia, PA 19122, USA}
\begin{abstract}

The newly developed "strongly constrained and appropriately normed" (SCAN) meta-generalized-gradient approximation (meta-GGA) can generally improve over the non-empirical Perdew-Burke-Ernzerhof (PBE) GGA not only for strong chemical bonding, but also for the intermediate-range van der Waals (vdW) interaction. However, the long-range vdW interaction is still missing. To remedy this, we propose here pairing SCAN with the non-local correlation part from the rVV10 vdW density functional, with only two empirical parameters. The resulting SCAN+rVV10 yields excellent geometric and energetic results not only for molecular systems, but also for solids and layered-structure materials, as well as the adsorption of benzene on coinage metal surfaces. Especially, SCAN+rVV10 outperforms all current methods with comparable computational efficiencies, accurately reproducing the three most fundamental parameters---the inter-layer binding energies, inter-, and intra-layer lattice constants---for 28 layered-structure materials. Hence, we have achieved with SCAN+rVV10 a promising vdW density functional for general geometries, with minimal empiricism.

\end{abstract}


\maketitle


In 2004, graphene, the first two-dimensional (2D) material, was experimentally realized \cite{Novoselov2004}, triggering the renaissance of layered materials in both condensed matter physics and materials science. In the same year, the first general-geometry van der Waals (vdW) density functional was devised \cite{Dion2004}, followed by a series of efforts \cite{Lee2010d, Cooper2010, Klimes2010a, Klimes2011, Hamada2014, Vydrov2009, Vydrov2010} to include the vdW interaction within the framework of Kohn--Sham density functional theory (DFT) \cite{Kohn1965}, the current mainstream first-principles approach. Layered materials including graphene, $hex$-BN, transition-metal dichalcogenides, black phosphorus, etc., have demonstrated various new concepts and potential technical applications \cite{Wang2012e, Chhowalla2013, Xu2013a, Butler2013a}. By contrast, despite a lot of progress, people are still struggling to find a vdW density functional with an acceptable accuracy for layered materials \cite{Bjorkman2012a, Bjorkman2012, Bjorkman2012b, Bjorkman2014}. For example, none of the density functionals tested in Refs.\,\onlinecite{Bjorkman2012} and \onlinecite{Bjorkman2014} can simultaneously reproduce the high-level theoretical inter-layer binding energy, as well as the experimental inter- and intra-layer lattice constants with a satisfying accuracy for 28 layered materials. Accurate description of the layer--layer vdW interaction is self-evidently important for studying the physical properties of layered materials, especially the evolution of property as a function of configuration. To explore the interactions between the layered materials and the environment such as substrates and molecular absorbates further requires accurate treatment of the vdW and other kinds of chemical bonding on the same footing. In short, a better versatile vdW density functional has been long-awaited and is urgently needed. 

The difficulty to describe the layered materials arises from the nonlocal and long-range nature of the vdW interaction, as well as the coexistence of the weak vdW bonding and much stronger chemical bonding in layered materials. The vdW interaction originates from dynamic electron correlations, causing a net attraction between fragments \cite{Berland2015}. It has negligible effects in most bulk systems, becomes noticeable in some bulk solids like soft alkali metals \cite{Tao2010}, and is significant in sparse matter including molecular complexes, molecular crystals, layered materials, and surface-adsorbate systems, as well as the so-called ``soft matter''. Fully accounting for the vdW interaction is achievable by high-level methods such as quantum Monte-Carlo (QMC) \cite{Foulkes2001}, coupled-cluster singles and doubles with perturbative triples [CCSD(T)] \cite{Raghavachari1989}, and adiabatic-connection fluctuation-dissipation theorem within the random-phase approximation (RPA) \cite{Eshuis2012}, which are however only feasible for limited-size systems because of high computational cost.

Aiming for a better efficiency, various approaches have been proposed within the framework of DFT. The popular ones include the DFT+D series \cite{Grimme2006, Grimme2010a}, Tkatchenko-Scheffler (TS) methods \cite{Tkatchenko2009,Tkatchenko2012,Liu2015j}, the Rutgers--Chalmers vdW-DF family \cite{Dion2004, Lee2010d, Cooper2010, Klimes2010a, Klimes2011, Hamada2014}, as well as the VV10 \cite{Vydrov2010} and the rVV10 \cite{Sabatini2013} methods. The latter three take only the electron density and its first derivative as inputs, and hence are conceptually applicable to any chemical environment with general geometry. In these methods, the total $xc$ energy consists of the local/semilocal $xc$ and nonlocal correlation components: $E_{xc} = E_{xc}^0 + E_c^{nl}$. (For the VV10 and rVV10, an extra term $\beta N$ with N being the total electron number is required such that $E_{xc}^{nl} + \beta N = 0$ for the uniform electron gas \cite{Vydrov2010}.) Comparing with the Rutgers-Chalmers vdW-DFs, the VV10 and rVV10 methods are more flexible in form thanks to two parameters $C$ and $b$: $b$ allows for any $E_{xc}^0$, while $C$ (independently of $b$) is chosen to yield accurate asymptotic vdW interactions $-C_6/R^6$ for molecules. The original VV10 \cite{Vydrov2010} takes $E_{xc}^0 = E_x^{rPW86}+E_c^{PBE}$ \cite{Murray2009a, Perdew1996}, $C=0.0093$, and $b=5.9$. The rVV10 method differs from VV10 only by a slightly \textit{revised} form together with a different $b=6.3$ \cite{Sabatini2013}. With such settings, VV10 and rVV10 describe the tested molecular systems very well \cite{Vydrov2010, Sabatini2013}, and reproduce remarkably well the experimental intra- and inter-layer lattice constants for 28 layered materials \cite{Bjorkman2012}. However, VV10 and rVV10 overestimate the inter-layer binding energies with respect to the RPA results by around 40\%, although \textit{in a preferred systematic way} \cite{Bjorkman2012}. The inter--layer binding energies from VV10 and rVV10 can be improved by a refitted $b$ parameter ($b=9.15$ for both), however at the undesired loss of accuracy for the lattice constants of layered materials, and for other systems like molecular complexes \cite{Bjorkman2012b}.

Here, we focus on the other control knob of the VV10 and rVV10 methods, $E_{xc}^0$, introducing the meta-GGA for the first time into the field of nonlocal vdW density functionals. Recently, the ``strongly constrained and appropriately normed'' (SCAN) meta-GGA \cite{Sun2015a} was constructed based on \textit{all 17 known exact constraints} appropriate to a semilocal functional, and a set of ``appropriate norms'' for which a semilocal function can be exact or nearly exact, like slowly-varying electron densities and compressed Ar$_2$. Presenting remarkable accuracy and versatility for the structural and energetic properties in strongly bonded systems, SCAN also gets the intermediate-range vdW interaction about right, which was otherwise severely overestimated by the local density approximation (LDA) and severely underestimated by other nonempirical semilocal functionals. (The intermediate range is roughly the distance between nearest-neighbor atoms at equilibrium) This has been demonstrated by the improved interaction energies for the S22 data set \cite{Sun2015a}, and better geometric and energetic properties of ice in different phases \cite{Sununpub}. With SCAN for $E_{xc}^{0}$ and appropriate parametrization, the VV10 or rVV10 $E_c^{nl}$ will take effect only for the long-range vdW interaction, and keep the advantages of SCAN intact. In this study, we demonstrate this concept with the rVV10 method, which allows for a fast implementation \cite{Roman-Perez2009}. The resultant SCAN+rVV10, with a single $b$ parameter and the fixed $C$, turns out to be a successful versatile vdW density functional, with a high accuracy for different systems including molecular complexes, bulk solids, benzene adsorbed on coinage metal surfaces, and the more challenging layered materials.

\section{Results}

\subsection{Parameters in SCAN+rVV10}

The rVV10 nonlocal correlation functional takes the same standard form as the Rutgers-Chalmers vdW-DFs,
\begin{equation}
\label{eq:ec_nl}
E_c^{nl} = \frac{\hbar}{2}\int\!\!\!\!\int\!\!d\boldsymbol{r}d\boldsymbol{r^{\prime}} n(\boldsymbol{r}) \Phi(\boldsymbol{r}, \boldsymbol{r^{\prime}}) n(\boldsymbol{r^{\prime}}),
\end{equation}
where $n(\boldsymbol{r})$ is the electron density, and $\Phi(\boldsymbol{r}, \boldsymbol{r^{\prime}})$ is the kernel describing the density-density interactions. Two empirical parameters $C$ and $b$ appear in the kernel: $C$ chosen for accurate $-C_6/R^6$ vdW interactions between molecules at large separation $R$, and $b$ controlling the damping of $E_c^{nl}$ at short range. For a semilocal $E_{xc}^0$, $C=0.0093$ was recommended \cite{Vydrov2010}, and the $b$ parameter was determined as $5.9$ and $6.3$ by fitting to the interaction energies of the S22 set \cite{Podeszwa2010,Takatani2010} for the original VV10 and rVV10  \cite{Vydrov2010, Sabatini2013}. Bj\"orkman \textit{et\,al} further proposed $b=9.15$ for VV10 by fitting to the binding energies of 26 layered materials, and we got the same number for rVV10. Increasing $C$ or $b$ generally results in smaller vdW correction. (In the following, we use ``rVV10'' to specifically denote the original rVV10 density functional with $E_{xc}^0 = E_x^{rPW86}+E_c^{PBE}$, $C=0.0093$, and $b=6.3$ \cite{Sabatini2013}.)

\begin{figure}[!htbp]                                                                                                   
\begin{center}
\caption{(Color online) The binding curves for Ar and Kr dimers from PBE, rVV10, SCAN, and SCAN+rVV10 compared to CCSD(T) curves \cite{Patkowski2005,Slavicek2003} as the reference (REF). } 
\label{fig:dimer}
\includegraphics[width=3.4in]{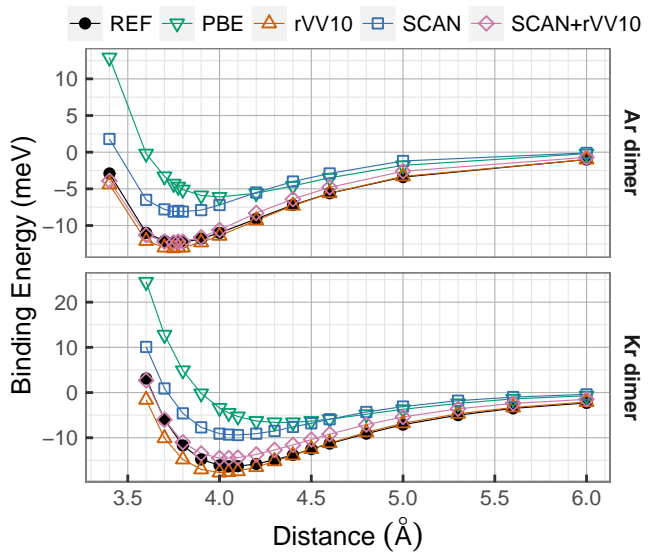}
\end{center}
\end{figure}

Here we determine the $b$ parameter by fitting to eleven data points around the equilibrium bond length from the CCSD(T) binding curve of the Ar dimer \cite{Patkowski2005}. (One point with a binding energy close to zero was excluded.) Note that three data points on the repulsive wall have been used as an ``appropriate norm'' for constructing SCAN \cite{Sun2015a}. With this choice, the more diverse S22 data set is reserved for benchmarking in the following, and the computational cost for the fitting is largely reduced. In the end, we obtain a mean absolute relative error of $2.6\%$ for these selected data points with $b=15.7$. The binding curve for Kr$_2$ was also computed with the same parametrization. In Fig.\,\ref{fig:dimer}, the two binding curves obtained from PBE, rVV10, SCAN, and SCAN+rVV10 are compared with the CCSD(T) results \cite{Patkowski2005, Slavicek2003}. Varying the value of $C$ does not noticeably improve the binding curve, and hence we stick to the recommended value of $0.0093$. 

\subsection{SCAN+rVV10 for molecular complexes}

\begin{figure}[!htbp]                                                                                                   
\begin{center}
\caption{(Color online) Box-plots \cite{boxplot} for the absolute relative errors and relative errors of the interaction energies from PBE, rVV10, vdW-DF2 (numerical results from Ref.\,\onlinecite{Vydrov2010}), SCAN, and SCAN+rVV10 with respect to the CCSD(T) results \cite{Podeszwa2010,Takatani2010}, for the molecular dimers in the S22 dataset. The PBE errors have been scaled down by a factor of 3 for clarity. The shapes inside the box denote the mean values.}
\label{fig:S22}
\includegraphics[width=3.4in]{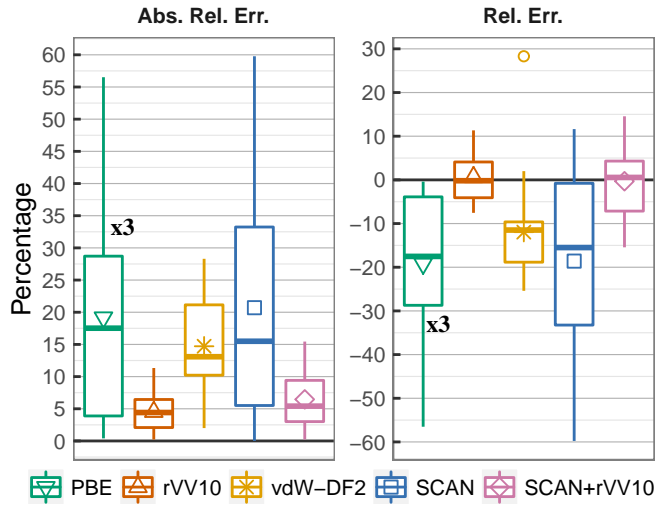}
\end{center}
\end{figure}

We first evaluate the performance of SCAN+rVV10 for molecular complexes by calculating the interaction energies for the S22 set which includes seven hydrogen-bonded, eight dispersion-bound, and seven mixed complexes. In Fig.\,\ref{fig:S22}, we compare the absolute relative errors and relative errors for the interaction energies from PBE, rVV10, vdW-DF2, SCAN, and SCAN+rVV10, with the CCSD(T) results \cite{Podeszwa2010, Takatani2010} as the reference. One striking point from Fig.\,\ref{fig:S22} (and also Fig.\,\ref{fig:dimer}) is the noticeable improvement over PBE by SCAN for the molecular systems. With the kinetic energy density as an extra input, meta-GGA can recognize covalent, metallic, and weak bonds \cite{Sun2013b}, so it is conceptually not surprising that some meta-GGAs can include---to some extent---the intermediate-range vdW interactions. For example, the M06-L \cite{Zhao2006} has a performance comparable to that of SCAN for the S22 data set. However, the M06-L used molecular systems in its construction, while SCAN is not fitted to any bonded system. Constrained by the 17 exact constraints, and guided by a set of appropriate norms, the resulting mathematical form of SCAN achieves excellent control of error cancellation between semilocal exchange and correlation, not only for covalent bonds as LDA and GGA do, but also for the intermediate-range vdW interaction, although the resulting binding still comes from the exchange part instead of the correlation \cite{Berland2013a, Zhang1997}. 
With such improvement for the intermediate-range vdW interaction, SCAN requires a smaller correction from $E_c^{nl}$ and therefore a larger $b$ parameter. Indeed SCAN+rVV10 with $b$ as large as $15.7$ approaches the accuracy of the original rVV10 for these molecular systems, better than the vdW-DF2 (numerical results from Ref.\,\onlinecite{Vydrov2010}) which underestimates the interaction energies for S22 by about $15\%$ on average. The large $b$ parameter in turn keeps intact the excellent capability of SCAN to describe non-vdW systems, which is the subject of the following section.

\subsection{SCAN+rVV10 for bulk solids}

\begin{figure}[!htbp]                                                                                                   
\begin{center}
\caption{(Color online) Box-plots for the absolute relative errors and relative errors of the atomization energies $E_a$, and the lattice equilibrium volumes $V_0$, from RPA, PBE, SCAN, and SCAN+rVV10 for 50 solids, with respect to the experimental values. The RPA, PBE, and experimental values are from Refs.\,\onlinecite{Harl2010a} and \onlinecite{Schimka2013a}. }
\label{fig:solid}
\includegraphics[width=3.4in]{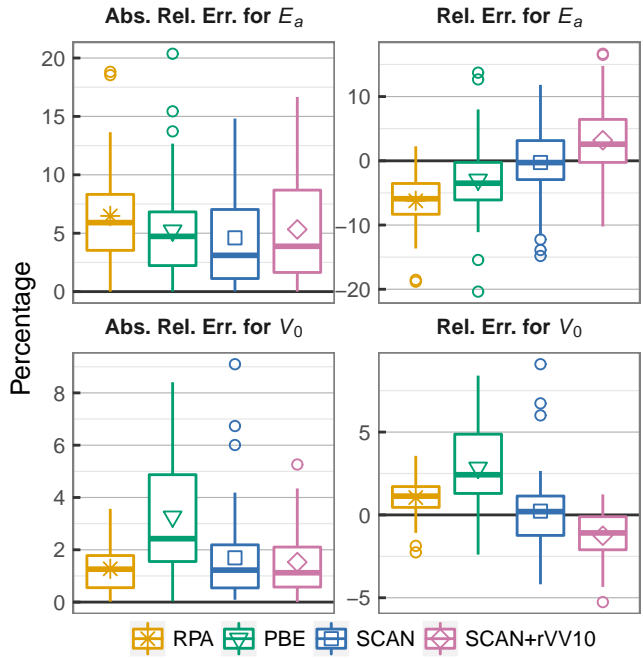}
\end{center}
\end{figure}

To benchmark the performance of SCAN+rVV10 for systems where vdW has little or only a slight effect, we compiled a set of 50 solid systems studied previously in Refs.\,\onlinecite{Harl2010a} and \onlinecite{Schimka2013a}, which includes ($\rmnum{1}$) 13 group--$\Rmnum{4}$ and $\Rmnum{3}$--$\Rmnum{5}$ semiconductors, ($\rmnum{2}$) 5 insulators, ($\rmnum{3}$) 8 main-group metals, ($\rmnum{4}$) 3 ferromagnetic transition metals Fe, Co, and Ni, and ($\rmnum{5}$) 21 other transition metals for which non-spin-polarized calculations were performed. For this set of solids, atomization energies and lattice volumes from RPA, and the corresponding experimental values after the zero-point correction, are available \cite{Harl2010a, Schimka2013a}. In Fig.\,\ref{fig:solid}, the comparison between the RPA, PBE, SCAN, and SCAN+rVV10 is given. For the atomization energy, both SCAN+rVV10 and SCAN are only slightly better than PBE and RPA. (However, atomization energy may not be a good choice to assess a semilocal functional \cite{Perdew2015}.) For the lattice volume, we found that SCAN+rVV10 is essentially as good as RPA, behaves similarly to SCAN for most solids, and is much better than the PBE functional which overestimates with the mean absolute relative error about 3\%. SCAN overestimates the volume by 6--10\% for K, Rb, and Cs (the three outliers), because of the long-range vdW effect in these soft alkali metals. This vdW effect originates from the long-range attraction between semicore $p$ orbitals \cite{Tao2010}, which SCAN fails to describe. Not unexpectedly SCAN+rVV10 properly improves the volumes for these three, and does not skew the distribution for the other solids as desired.

\subsection{SCAN+rVV10 for benzene adsorbed on coinage metal surfaces}

The outstanding performance of SCAN+rVV10 for the molecular and solid systems above is encouraging. To be a successful general-geometry density functional, SCAN+rVV10 should pass more stringent tests, where both strong local atomic bonds and weak vdW interactions need to be well described. Here we consider the widely studied benzene ring adsorbed on the (111) surface of coinage metals \cite{Bilic2006a, Toyoda2009, Liu2013x, Yildirim2013, Reckien2014, Carter2014}. This is also one of the most common systems for benchmarking a vdW functional's capability to describe simultaneously the vdW bonding and the metallic bonding. At first we want to emphasize that the lattice constants from SCAN+rVV10 are 3.544, 4.058, and 4.073\,\AA\ for Cu, Ag, and Au, in excellent agreement with the experimental values of 3.595, 4.062, and 4.062\,\AA. This is much better than the vdW-DF1 methods with different exchange density functionals \cite{Yildirim2013}. In Table \ref{tab:bz}, we compare the calculated adsorption energy $E_{ad}$, and distance $\Delta z$ between the benzene and the surface from SCAN and SCAN+rVV10, with available experimental values \cite{Liu2015j, Campbell2012, Xi1994, Zhou1990, Syomin2001, Ferrighi2011a}. SCAN+rVV10 agrees well with experiment for $E_{ad}$, while SCAN systematically underestimates by about $0.4$ eV. The experimental values for the distance between benzene and the surfaces were measured for Ag \cite{Liu2015j} and estimated for Au \cite{Ferrighi2011a}, both of which are close to our SCAN+rVV10 results. Considering the chemical trend of the ionic radii, we are confident in our SCAN+rVV10 prediction for $\Delta z$ in the case of Cu. The results from other methods, including PBE, PBE+vdW, PBE+vdW$^{surf}$, MP2, vdW-DF1, DFT+D, and M06-L, etc., have been compiled in Ref.\,[\onlinecite{Liu2013x}]. Unfortunately, none of them predicts simultaneously the correct $E_{ad}$ and $\Delta z$, with the best results from PBE+vdW$^{surf}$ \cite{Liu2013x}, which still overestimates the adsorption for the case of Cu and apparently cannot apply to general geometries. The recent rev-vdW-DF2 \cite{Hamada2014} is close to SCAN+rVV10 based on its results for benzene on Cu(111), but more thorough calculations are needed for confirmation. Since the adsorption of organic molecules on metallic surfaces plays an important role in catalysis, molecular sensors and switches, etc., SCAN+rVV10 will be a very helpful tool in these fields. 

\begin{table}[!htbp]
\caption{Adsorption energy $E_{ad}$ and distance $\Delta z$ between benzene and the (111) surface of Cu, Ag, and Au from SCAN and SCAN+rVV10, compared with experimental results \cite{Liu2015j,Campbell2012, Xi1994, Zhou1990, Syomin2001, Ferrighi2011a} when available. The data for the lowest-energy hcp30$^{\circ}$ configuration \cite{Liu2013x} is shown.}
\label{tab:bz}
\begin{ruledtabular}
\begin{tabular}{lcccccc}
	& \multicolumn{2}{c}{SCAN}  &  \multicolumn{2}{c}{SCAN+rVV10} & \multicolumn{2}{c}{Experiment}\\
\hline
	& $E_{ad}$ (eV) & $\Delta z$ (\AA) & $E_{ad}$ (eV) & $\Delta z$ (\AA) & $E_{ad}$ (eV) & $\Delta z$ (\AA) \\
Cu	&	$0.36$ & $3.0$ 	& $0.74$ & $2.93$ & $0.71$ & N/A \\
Ag 	& 	$0.34$ & $3.2$	& $0.68$ & $3.02$ & $0.68$ & 3.04 \\
Au	& 	$0.37$ & $3.2$	& $0.73$ & $3.07$ & $0.76$ & $\sim3$ \\
\end{tabular}
\end{ruledtabular}
\end{table}

\subsection{SCAN+rVV10 for layered materials}

\begin{figure}[!htbp]                                                                                                   
\begin{center}
\caption{(Color online) Box-plots for the absolute relative errors and relative errors of the inter-layer binding energies, inter- and intra-layer lattice constants ($c$ and $a$) from rev-vdW-DF2, SCAN, and SCAN+rVV10, for 28 layered materials. The reference values are from RPA for the binding energy, and from experiment for the lattice constants \cite{Bjorkman2012, Bjorkman2014}.}
\label{fig:2D}
\includegraphics[width=3.4in]{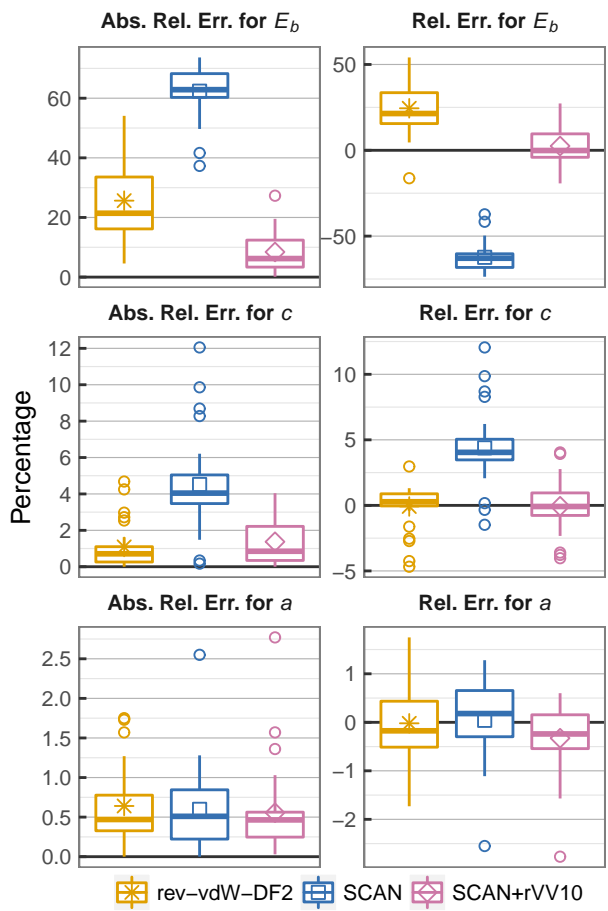}
\end{center}
\end{figure}

Layered materials have become one of the main arenas for condensed matter physics as well as materials science during the last decade \cite{Wang2012e, Chhowalla2013, Xu2013a, Butler2013a}. When embarking on first-principles exploration of known or unknown 2D materials \cite{Bjorkman2012, Inoshita2014}, three most fundamental quantities have to be correctly predicted: the inter-layer binding energy ($E_b$), the inter-layer lattice constant ($c$), and the intra-layer lattice constant ($a$). The first two are mainly determined by vdW interactions, and the last one by the stronger covalent or ionic bonding. Bj\"orkman et al.\,\cite{Bjorkman2012, Bjorkman2014} have tested the performance for a series of vdW density functionals, using $E_b$ from RPA calculation and experimental lattice constants as the references for 28 hexagonal layered materials. One conclusion is that no single functional yet can yield a mean absolute relative error $<10\%$ for $E_b$, $<2\%$ for $c$, and $<1\%$ for $a$. Among the tested functionals, the recently proposed rev-vdW-DF2 \cite{Hamada2014} was found to be the best for the lattice constants, together with a relatively small mean absolute relative error of 16\% for $E_b$. In Fig.\,\ref{fig:2D}, we compare the performance of this functional, SCAN, and SCAN+rVV10 for 28 layered materials for which the RPA $E_b$ are available \cite{Bjorkman2012, Bjorkman2014}. (Here we found the mean absolute relative error of $E_b$ from rev-vdW-DF2 close to 26\%, which may arise from the different implementation.) As expected, SCAN predicts an accurate intra-layer lattice constant for which vdW has little effect (there is one outlier from VTe$_2$ with $a$ underestimated by 3\%, which may be due to the self interaction error). However SCAN does strongly underbind along the $c$ direction, as illustrated by the systematically $\sim60\%$ underestimated $E_b$, and the largely scattered absolute relative errors of $c$. With the $b$ parameter determined by the Ar$_2$ binding curve, the mean absolute relative error for $c$ from SCAN+rVV10 drastically reduces to 1.4\%, comparable with that from rev-vdW-DF2, and the mean absolute relative error for $E_b$ dramatically reduces to 8\%, only about 1/3 of that from rev-vdW-DF2. As expected, the long-range rVV10 correction is much more important for the interlayer binding energy than it is for the S22 binding energies of small molecular complexes. Meanwhile, the mean absolute relative error for $a$ is as small as $0.6\%$, the same as those from SCAN and rev-vdW-DF2. Hence, SCAN+rVV10 becomes the first functional reducing the mean absolute relative error $<10\%$ for $E_b$, $<2\%$ for $c$, and $<1\%$ for $a$. This significant success suggests that SCAN+rVV10 can be the method of choice for layered materials, especially in the case of large-scale computation.

\section{Discussion}

The above benchmarking calculations have established SCAN+rVV10 as a successful versatile vdW density functional, especially in the near-equilibrium cases. The success builds on three facts: the high quality of SCAN as a semilocal density functional for even intermediate-range vdW interactions, the high quality of the nonlocal correlation functional in rVV10 for long-range vdW interactions, and flexibility in the rVV10 method to adjust the short-range damping via the $b$ parameter. 
We would expect similar success from a combination of SCAN with other long-range vdW corrections that have such flexibility: the DFT+D \cite{Grimme2006, Grimme2010a}, and Tkatchenko-Scheffler (TS) methods \cite{Tkatchenko2009, Tkatchenko2012, Liu2015j}. In these methods, \textit{ad hoc} pairwise inter-atomic terms are added to the DFT potentials, with explicit usage of $C_6$ parameters and a damping function. 
These methods are in principle ion- or atom-based, instead of electron density-based, and hence may not be applied as generally as the rVV10, VV10 and Rutgers--Chalmers vdW-DF methods \cite{Dion2004, Lee2010d, Berland2015}. However, incorporating SCAN within these methods is of special interest. For example, incorporating SCAN within the DFT+D methods, one can readily include high-order $C_8$, $C_{10}$ terms.

Our version of rVV10 is fitted to systems with fundamental energy gaps (Ar$_2$ and pairs of well--separated 
molecules), but it performs just as well for the gapless systems studied here (the coinage metals interacting with benzene, and the layered material graphite). We would not however expect rVV10 to be accurate for the long-range vdW asymptotics in systems of small or zero gap with more challenging geometries, such as pairs of large fullerene molecules \cite{Ruzsinszky2012, Gobre2013d}. While taking the electron density distribution into account \cite{Tkatchenko2009}, the more recent TS methods \cite{Tkatchenko2012, Liu2015j} further include 
screening effects and many-body dispersion, which  can be important for vdW asymptotics \cite{Gobre2013d}. Even when SCAN+rVV10 is not accurate for the long-range vdW coefficient $C_6$, it could still be accurate for near-equilibrium binding. This appears to be so for the zero-gap layered material graphite, where the interlayer energy for large interlayer separation $D$ is correctly $\sim D^{-3}$ in RPA \cite{Dobson2006}, but $\sim D^{-4}$ in pairwise-interaction models including SCAN+rVV10.

The Rutgers--Chalmers vdW-DF family \cite{Berland2015}, which inspired the VV10 and rVV10 methods, currently sticks to the LDA correlation in $E_{xc}^0$. A lot of effort has been spent on designing a better semilocal exchange functional \cite{Dion2004, Lee2010d, Cooper2010, Klimes2010a, Klimes2011, Hamada2014}, which ideally includes no vdW effects. We cannot expect improvement of the family of the Rutgers--Chalmers methods by directly incorporating SCAN, either only in the exchange part or the whole semilocal exchange correlation energy. However, other meta-GGA’s could provide better options for the semilocal part in this family.

\section{Conclusion}

In conclusion, we have devised a promising vdW density functional SCAN+rVV10 based on the SCAN meta-GGA. It works for general geometries as exemplified by its excellent performances for the molecular complexes, solids, benzene adsorption on coinage metal surfaces, and layered materials. SCAN+rVV10 achieves an accuracy comparable to that of higher-level methods like RPA and CCSD(T) for various situations benchmarked here, with a computational cost reduced by several orders of magnitude. SCAN+rVV10 outperforms other currently available methods with comparable computational efficiency for the energetics and structures of layered materials, making it the right choice for the computational study of 2D layered materials. 

\section{Methods}

All calculations in this study were performed with the projector augmented wave (PAW) method \cite{Blochl1994b} as implemented in the VASP code \cite{Kresse1994, Kresse1996, Kresse1999a}. The PAW Perdew-Burke-Ernzerhof (PBE)  \cite{Perdew1996} pseudopotentials (version .52) recommended by the VASP developers were employed. The rVV10 nonlocal correlation functional has been implemented within the VASP code based on the vdW-DF implementation by Klime\v s \textit{et al}\,\cite{Klimes2010a}, and has been benchmarked by comparing with the original rVV10 implementation \cite{Sabatini2013} within the Quantum-Espresso code. 

To calculate the binding curve of Ar$_2$, the dimer and the atom were put in a cubic supercell with a length of 25\,\AA, an energy cutoff for the plane wave basis as high as 1200 eV was chosen to ensure high quality for the fitting, and the single $\Gamma$ point was used for the Brillouin zone sampling. 

To calculate the interaction energies of the S22 data set \cite{Podeszwa2010,Takatani2010}, the molecular dimer or complex was put into an orthogonal supercell, where the periodic images were separated with a 20-\AA-thick vacuum slab along all three directions to avoid interaction between them. An energy cutoff of 900 eV was chosen. The same simulation cell, and the same computational parameters were used for the components of the molecular dimer or complex to minimize errors. For all molecular calculations, the single $\Gamma$ point was used for the Brillouin zone sampling. 

For the 50 solid systems, the initial atomic structures for the solids, and the spin configuration for the isolated atoms, were chosen the same as those in Ref.\,[\onlinecite{Harl2010a}] and Ref.\,[\onlinecite{Schimka2013a}]. 800 eV was chosen for the basis energy cutoff. The isolated atom was simulated with a $14\times15\times16$ \AA$^3$ orthogonal cell. For a solid, a $\Gamma$-centered Monkhorst-Pack \cite{Monkhorst1976b} $k$-mesh corresponding to at least $4000$ per reciprocal atom was used for the Brillouin zone integration. Both the lattice vectors, and internal atomic coordinates have been relaxed until the residual atomic force is less than $0.01$\, eV/\AA.

The initial atomic models for the benzene molecule on Cu, Ag, and Au (111) surfaces were constructed based on those in Ref.\,[\onlinecite{Bilic2006a}]. The metallic surfaces were modelled with five $3\times3$ atomic layers, with the bottom three layers fixed to their bulk coordinates. A 20\,\AA-thick vacuum slab was inserted between the metal slab and its periodic images. During the relaxation, the lattice vectors were fixed, with only the internal atomic coordinates of the top two metal atomic layers and the benzene molecule optimized. The energy cutoff was 600 eV, and the $\Gamma$-centered $4\times4\times1$ Monkhorst-Pack \cite{Monkhorst1976b} $k$-meshes were used. For SCAN, since the binding curve is very shallow, we further calculated the adsorption energies at distances $\Delta z$ from 2.8 to 3.6 \AA\ with a step of 0.1 \AA. This is the reason why we only keep one decimal digit for the SCAN $\Delta z$ in Table I of the main text. The same was done for SCAN+rVV10, but we found SCAN+rVV10 is able to find the minima with a standard atomic relaxation.

For the layered-structure materials, the initial atomic structures were taken from Ref.\,[\onlinecite{Bjorkman2012}] and Ref.\,[\onlinecite{Bjorkman2014}]. The energy cutoff for the plane wave basis was 800 eV, and the $\Gamma$-centered $12\times12\times6$ and $12\times12\times1$ Monkhorst-Pack \cite{Monkhorst1976b} $k$-meshes were used for the bulk form and monolayer, respectively. A 30\,\AA-thick vacuum slab was utilized to model the monolayers. The internal coordinates were always fully relaxed. To determine the lattice vectors of the bulk structures, we further relaxed all three lattice vectors. To determine the binding energy, we fixed the intra-layer lattice constants to their experimental values, and relaxed the inter-layer lattice constants only for the bulk structures, as in the RPA calculations \cite{Bjorkman2012, Bjorkman2014}. The relaxation criteria were the same as those for the 50 solids.   

\ 

\begin{acknowledgments}
This work was supported as part of the Center for the Computational Design of Functional Layered Materials, an Energy Frontier Research Center funded by the U.S.\,Department of Energy, Office of Science, Basic Energy Sciences under Award \#.\,DE-SC0012575. This research used resources of National Energy Research Scientific Computing Center (NERSC), a DOE Office of Science User Facility supported by the Office of Science of the U.S.\,Department of Energy.
\end{acknowledgments}

%
\end{document}